\documentclass[conference]{IEEEtran}
\IEEEoverridecommandlockouts
\usepackage{cite}
\usepackage{amsmath,amssymb,amsfonts}
\usepackage{algorithmic}
\usepackage{graphicx}
\usepackage{textcomp}
\usepackage{url}
\usepackage{mdframed}
\usepackage{booktabs}
\usepackage{xcolor}
\usepackage{csquotes}
\usepackage[]{hyperref}
\def\BibTeX{{\rm B\kern-.05em{\sc i\kern-.025em b}\kern-.08em
    T\kern-.1667em\lower.7ex\hbox{E}\kern-.125emX}}
\newmdenv[backgroundcolor=yellow!10,
            leftline=false,
            rightline=false,
            bottomline=false,
            linewidth=2pt,
            linecolor=blue]{myframe}
\newcommand{\visionhead}[1]{
\begin{myframe}
\noindent {\textbf{#1}} 
\end{myframe}

}

\begin{document}

\title{Technical Debt Management: The Road Ahead for Successful Software Delivery\\
}

\makeatletter
\newcommand{\linebreakand}{%
  \end{@IEEEauthorhalign}
  \hfill\mbox{}\par
  \mbox{}\hfill\begin{@IEEEauthorhalign}
}
\makeatother

\author{\IEEEauthorblockN{Paris Avgeriou}
\IEEEauthorblockA{\textit{Dept. of Computing Science} \\
\textit{University of Groningen}\\
Groningen, The Netherlands \\
p.avgeriou@rug.nl}
\and
\IEEEauthorblockN{Ipek Ozkaya}
\IEEEauthorblockA{\textit{Software Engineering Institute} \\
\textit{Carnegie Mellon University}\\
Pittsburgh, PA, USA \\
ozkaya@sei.cmu.edu}
\and
\IEEEauthorblockN{Alexander Chatzigeorgiou}
\IEEEauthorblockA{\textit{Dept. of Applied Informatics} \\
\textit{University of Macedonia}\\
Thessaloniki, Greece \\
achat@uom.edu.gr}
\linebreakand
\IEEEauthorblockN{Marcus Ciolkowski}
\IEEEauthorblockA{\textit{QAware GmbH} \\
München, Germany \\
marcus.ciolkowski@qaware.de}
\and
\IEEEauthorblockN{Neil A. Ernst }
\IEEEauthorblockA{\textit{Department of Computer Science} \\
\textit{University of Victoria}\\
Victoria, Canada \\
nernst@uvic.ca}
\and
\IEEEauthorblockN{Ronald J. Koontz}
\IEEEauthorblockA{\textit{The Boeing Company} \\
Mesa, AZ, USA \\
ron.j.koontz@boeing.com}
\linebreakand
\IEEEauthorblockN{Eltjo Poort}
\IEEEauthorblockA{\textit{CGI} \\
Rotterdam, The Netherlands \\
eltjo.poort@cgi.com}
\and
\IEEEauthorblockN{Forrest Shull}
\IEEEauthorblockA{\textit{Software Engineering Institute} \\
\textit{Carnegie Mellon University}\\
Pittsburgh, PA, USA \\
fjshull@sei.cmu.edu}
}

\maketitle

\begin{abstract}
Technical Debt, considered by many to be the ‘silent killer’ of software projects, has undeniably become part of the everyday vocabulary of software engineers. We know it compromises the internal quality of a system, either deliberately or inadvertently. We understand Technical Debt is not all derogatory, often serving the purpose of expediency. But, it is associated with a clear risk, especially for large and complex systems with extended service life: if we do not properly manage Technical Debt, it threatens to “bankrupt” those systems. Software engineers and organizations that develop software-intensive systems are facing an increasingly more dire future state of those systems if they do not start incorporating Technical Debt management into their day to day practice. But how? What have the wins and losses of the past decade of research and practice in managing Technical Debt taught us and where should we focus next? In this paper, we examine the state of the art in both industry and research communities in managing Technical Debt; 
we subsequently distill the gaps in industrial practice and the research shortcomings, and synthesize them to define and articulate a vision for what Technical Debt management looks like five years hence.
\end{abstract}

\begin{IEEEkeywords}
technical debt, software maintenance and evolution
\end{IEEEkeywords}

\section{Introduction}
Imagine a scenario where a development team is under pressure to quickly add a high-value product feature and must choose between building the feature ‘first time right’ or building it ‘quick-and-easy’~\cite{poort23}. 
The development team will likely choose the `quick-and-easy' solution, with the intention of refactoring to the more robust solution later if one or more of the following conditions exist: 1) significant time pressure, 2) insufficient funding, 3) unavailability of the required engineering resources, 4) large feature size and/or feature complexity. 
This tradeoff decision can have far-reaching consequences, ranging from complex and hard to maintain code to reduced team velocity and frequent, unexpected rework. The `quick and easy' solution often multiplies and becomes permanent, thereby eroding the product’s architecture and limiting the ability to make future updates. The \textbf{Technical Debt} metaphor has proven to be a useful vehicle to help understand the behavioral and economic aspects of tradeoffs like this, where a team or product benefit in the short-term but suffer negative consequences in the long-term \cite{Cunningham1992}.

Software engineers perform these tradeoffs routinely, sometimes deliberately, other times inadvertently. On occasion, they have access to complete information; at times, they may need to rely on partial data. They could be aware of the uncertainties around the evidence they have, but in many cases they are not. 
At various intervals, competing social, financial, and strategic priorities influence the tradeoffs in unexpected ways.  Such technical and management decisions that imply a tradeoff among outcomes at different points in time (immediate benefit versus increased future maintenance costs) have been widely studied in economics. Intertemporal choices, where distant outcomes are often valued lower than short-range ones lead to \textit{temporal discounting} \cite{Soman2005}, a phenomenon which over and over leads to unmanageable Technical Debt. 

The tradeoff decision-making process is not unique to software; it is the nature of engineering. However, the ability of a single software engineer to quickly implement and deploy changes to an end product differentiates software engineering from other engineering disciplines, where such modifications are likely to occur far more gradually.

The concept of Technical Debt has been embraced by software engineers ever since Ward Cunningham coined the term in 1992 \cite{Cunningham1992}, as the idea effectively encapsulates multiple aspects of tradeoff decision-making in software development while accounting for both value and cost. In his original premise, Cunningham specifically emphasized the need to refactor code that is often written with only partial knowledge of the problem and domain, and without enough upfront design \cite{Cunningham1992}. Since then, the Technical Debt perspective has shaped the way both industry and academia think of shortcuts taken in software development, viewing them as a consequence of tradeoffs that expedite software release at the potential risk of higher cost for changes in the long term.

Since Ward Cunningham’s original proclamation, significant progress has been made in raising awareness of software engineers to understand and manage Technical Debt. There is now widespread consensus that managing Technical Debt should be treated as a core software engineering practice, applied continuously across the software development life cycle \cite{Kruchten2019}. Industry is increasingly incorporating Technical Debt management practices to their development processes \cite{Jaspan2023, PaulischMTD2016}. Many software quality tools now incorporate features to help software engineering teams visualize and triage Technical Debt issues within their code bases. Software project management and issue tracker tools have also experimented with including Technical Debt as one of the default labels, e.g. Planview \footnote{https://www.planview.com/}.

This advancement in industrial practice is paired with very vibrant research. The research community has produced a substantial body of knowledge on the topic of Technical Debt, especially in investigating the problem and its manifestation, understanding its urgency and impact, and proposing solutions to address it. This research is often performed in collaboration with industry, indicating its practical relevance but also the aligned interests of researchers and practitioners.

In this paper, we examine current progress made by industry and research communities and then lay out a path forward with some substantial shifts in theory and practice. We first clarify concepts related to Technical Debt in Section \ref{sec:whatis}. We then discuss the challenges and current state of industry practice regarding Technical Debt, through the lenses of four different representative organizations in Section \ref{sec:industry}. Subsequently we take a critical review of the latest research in managing Technical Debt, including the available tools and research data sets in Section \ref{sec:research}. Finally, we present a vision for effectively managing Technical Debt where researchers, tool vendors, and practitioners collaborate to close the gaps in Section \ref{sec:vision}.

\section{What is Technical Debt?}
\label{sec:whatis}

Like most popular terms in software engineering, Technical Debt has been defined time and again in both gray literature (especially blogs) and scientific literature. We simply rely on the definition that emerged from a Dagstuhl Seminar on this topic through consensus, and has stood the test of time \cite{DagstuhlAvgeriou2016}:

\emph{\begin{displayquote}
In software-intensive systems, Technical Debt is a collection of design or implementation constructs that are expedient in the short term but set up a technical context that can make future changes more costly or impossible. Technical Debt presents an actual or contingent liability whose impact is limited to internal system qualities, primarily maintainability and evolvability.
\end{displayquote}}

This definition covers the essential elements of Technical Debt, also anchored by the original intent of Cunningham, emphasizing \emph{“a little debt speeds development as long as it is paid back promptly with a rewrite”}:

\begin{itemize}
        \item It highlights that Technical Debt concerns not only implementation constructs (as often misunderstood or miscommunicated) but also earlier development artifacts that pertain to software requirements, architecture and design.
        \item  It reflects the tradeoff between short-term gain in certain business metrics (e.g., time to market or development cost) and long-term increase in cost of change (which can become prohibitive). This tradeoff has important implications for strategies to manage Technical Debt: some debt, consciously acquired, can be a good thing - for example, by delivering working software to end users faster so that they can provide feedback.
        \item  It captures that Technical Debt may directly affect current software development (e.g., by slowing development velocity) or it may contingently do so in the future. The latter is mostly unintentional Technical Debt: instead of the development team making design decisions deliberately, a change of circumstances can trigger a liability; e.g., a third-party library no longer being supported by its provider can increase maintenance costs.
        \item It emphasizes how Technical Debt directly affects design-time qualities, such as maintainability and evolvability of software-intensive systems. Within complex large-scale systems, Technical Debt can also indirectly affect other quality attributes, especially run-time qualities like security and reliability, in the sense that maintenance and evolution issues can in turn cause security vulnerabilities and bugs.
        \item It stresses a simple bottom line: \textbf{Technical Debt is about managing the impact of cost of change over time}.
\end{itemize}

\subsection{Technical Debt versus Other Debts and Issues}
\label{sec:tdvsother}

Both the scientific and gray literature are riddled with uses of the term ``debt'' in the context of software engineering activities. We often read about social, process, infrastructure, and management ``debt”, applying the concept to other software engineering activities, such as development-process execution, hardware-upgrade cycles, or people and team management. This often leads to confusion among both practitioners and researchers regarding the meaning of Technical Debt and its relation to other “debts”. We take no issue with such term usage; we see the potential utility in many of them and encourage the examination of their relation to Technical Debt. However, we do argue that these are separate notions from Technical Debt and need to be independently and clearly defined and established. We insist on this clear separation of terms for a simple reason: the power of the Technical Debt metaphor lies in it being clearly defined, scoped, and related to design or implementation constructs (including e.g. tests, build scripts, and algorithms). In contrast, if everything detrimental to software development is labeled as “debt”, then Technical Debt as a notion is diluted and loses its practical usage and essential emphasis on value-creation as one can no longer concretely identify, measure, or fix it.

Confusion also exists around the relationship between Technical Debt, security vulnerabilities, and defects (or bugs). While Technical Debt may also \textit{cause} security vulnerabilities (e.g., due to a deprecated library not being updated) or bugs (e.g., introducing unintentional defects when changing “smelly” or incomprehensible code), those vulnerabilities and bugs \textit{are not} actual Technical Debt. Conversely, the existence of bugs or vulnerabilities does not directly imply Technical Debt as their root cause.

In practical terms, there is one very simple criterion, a litmus test, to assess when something is Technical Debt or not: \textbf{if a design or implementation construct incurs interest when being changed, then it is indeed Technical Debt}. The notion of “interest” in this context is defined as \emph{“additional costs incurred by the project in the presence of Technical Debt”} \cite{Avgeriou2016}. In other words, if the existence of such a construct (e.g., a ‘God’ component or dependency cycles) causes an increase in the cost of change (e.g., implementing a feature takes longer) \cite{NordICSA2022}, then interest is being paid, and we can safely classify it as Technical Debt. As a simple example, when a dependency cycle among three components causes changes in any of these components to be costlier (compared to if the cycle was not there), then we clearly pay interest. Because of these dependency cycles, we may be incurring other additional costs as well: for example, increased operations costs due to run-time penalties. To prevent “watering down” the Technical Debt concept to anything that potentially reduces development budget, costs need to clearly map to changing design and implementation constructs; otherwise, they are not considered Technical Debt. 

\subsection{Technical Debt Research and Practice at An Inflection Point}

The interest of the research community on Technical Debt followed a 2010 Future of Software Engineering workshop paper  \cite{Brown2010} which put forward the following vision at the time:

\emph{The impact of this research, if it succeeds, will be improved software development productivity and quality. Software developers and managers will better reason about the liabilities and opportunities created by Technical Debt and make better decisions about managing them. Software engineers would understand the rationale that managers use in making such decisions. This will lead to improved software maintenance and, in the end, better software. Finally, software tool developers will have a new set of functions to support and new markets for their tools based on a coherent framing of the issues.}

Revisiting this vision statement, thirteen years later, there are three points to emphasize. First, improving software development productivity and quality, as well as enhancing tradeoff analysis and execution, may seem like goals specific to the development team. However, these principles also affect the \textit{users' ability to consistently receive value in future software updates}: addressing Technical Debt appropriately and at the right time enables teams to continually deploy new updates on appropriate cycles, while maintaining a justified confidence that the software performs as expected, is safe and secure, and can allow for future evolution. 

The second point relates to recent advances in software engineering tools, and their ability to address technical, evolutionary and organizational concerns in Technical Debt management: 

\begin{itemize}
\item On one hand, there is an accelerated increase in new automated features within development tools that aim to improve development efficiency and reliability for engineers. Emerging tools are increasingly investigating how to leverage the predictive capabilities of AI and language models trained on huge code bases. Their goal is mostly to enable developers to catch implementation errors as they occur, complete code correctly, and recommend/fix areas of code where straightforward refactoring opportunities exist.   

\item On the other hand, while the industry is driving the simplification of code generation with AI-augmented tools to develop vast new quantities of code and develop it fast, it risks resulting in greater potential for incurring unanticipated Technical Debt. This regression may be further worsened by the potential for a new generation of developers who are over reliant on AI support for basic programming tasks and do not develop sufficient expertise to assess quality, understand risks, explore tradoffs, and weigh decision consequences \cite{BirdCoPilotACM}.
\end{itemize}

The third point concerns the reality of industrial software engineering and specifically the significant amount of legacy software. In these heritage systems, no matter their pedigree, accumulated Technical Debt exists where software development teams do not have sufficient resources to address all existing issues. A large portion of that Technical Debt was introduced many years prior; it remains hidden and undocumented, and involves design or architecture issues that cannot be mined by analyzing source code. Any new tools and practices for managing Technical Debt must also deal with the harsh reality of brownfield development.  

In the next sections, our discussion of the current state of both industry practice and academic research and the challenges for practitioners and researchers today and for the future alike are motivated by these inflection points: the need to consistently provide value to end users under the pressure of accumulating Technical Debt, the evolving capabilities of automated tools and the ongoing brownfield development realities of existing software systems.   

\section{Current State of Industry Practice and Challenges}
\label{sec:industry}
Across the software industry, Technical Debt, by and large, is now part of the vernacular. The main reason for this terminology adoption is easier discussion of internal software quality: instead of explaining technical aspects like cohesion and coupling or design smells to managers, \textbf{Technical Debt attempts to put a dollar sign to these abstract qualities and helps justify investment of resources to pay back the debt}. More importantly, the monetization of said characteristics through Technical Debt expresses a clear risk: a debt running in the millions of dollars/euros is something development teams can urgently raise to their management. 

The software industry has evolved significantly since 1992, when Cunningham first introduced the term, and even since 2016, when a group of researchers, industry, and tool vendors at Dagstuhl established a formal Technical Debt definition (see Section~\ref{sec:whatis}). While the quest for “quantifying” Technical Debt continues, industry has also recognized that there is no one-size-fits-all tool or magic metric for Technical Debt, yet a toolbox is indeed needed to manage it. This realization is a big step forward for establishing an intentional Technical Debt management practice across the software industry. Furthermore, large software development organizations now also share their experience in managing Technical Debt and share techniques that may work better than others \cite{Jaspan2023, PaulischMTD2016, KazmanSoftServICSE2015}. 

As one example of such an industry organization, we consider Google, parent company now Alpha. Google has taken an empirical approach to understanding how Technical Debt manifests itself in their teams, publishing an engineering satisfaction survey since 2018 to understand how engineers may have been hindered by unnecessary complexity and Technical Debt \cite{Jaspan2023}. Survey outcomes have helped teams who are focused on developer productivity in Google understand the “somewhat” common areas of Technical Debt. In addition, Google explored 117 metrics as indicators of these common areas of Technical Debt (e.g., dependencies, code quality, migration, code degradation) and concluded that no single metric predicts reported Technical Debt. Their initial findings assert that there is no single generalizable metric to understand leading indicators of Technical Debt. This implies that teams must select and adjust metrics based on their context.

Technical Debt is not only crucially important in industry; its impact has increasingly been emphasized for any governments responsible for managing software at scale. For example, the significance of Technical Debt is sufficiently high to have been recognized by US Congress. In the FY22 National Defense Authorization Act (NDAA), which became public law in December 2021, Congress required a study of Technical Debt practices in the Department and recommendations for improvement \cite{NDAASec835, OzkayaNDAA}.

Based on our collective experience, there is no doubt that methods to manage Technical Debt will be increasingly relevant to industry. To help align targeted near-term and long-term research, we provide four different industry perspectives to exemplify current industry difficulties in dealing with Technical Debt: 1) dealing with Technical Debt in safety critical systems with insights from Boeing, 2) managing a large portfolio of projects which have accumulated Technical Debt as exemplified in government software management initiatives such as the US Department of Defense, 3) managing Technical Debt in an organization developing custom software solutions such as QAware, and 4) managing Technical Debt within the agile and business enterprise with experience from CGI.   These views provide exemplars of common problems and gaps between research ideas and tools and often overlooked realities of managing Technical Debt in the trenches. 

\subsection{Technical Debt and Safety Critical Systems}
Similar to the broader software industry, Technical Debt is a pervasive challenge within The Boeing Company \footnote {The Boeing Company is a global aerospace company with headquarters in Arlington, VA, United States of America.}, as it is prevalent in all software aspects and across all software life cycle stages. Military and commercial aircraft service life typically extends across several decades, where embedded avionics software typically undergoes multiple incremental upgrades and improvements.

Technical Debt symptoms which are not unique to Boeing and are usually visible to all stakeholders include:
\begin{itemize}
    \item Perpetually late deliveries
    \item System fragility
    \item Source code complexity
    \item Reduced productivity
    \item High maintenance cost
    \item Culture erosion
\end{itemize}

While the above symptoms can result from multiple root causes adjacent to Technical Debt, software-intensive systems impacted by several of these symptoms are likely struggling with excessive Technical Debt.  System fragility is often the result of early life cycle software architectural decisions that are expedient in the near-term (Technical Debt) that are not corrected and evolve into brittle/fragile mature systems that lack sufficient modularity and are not designed with well-defined interfaces.  For example, a change to one software component ripples across multiple components, thereby negatively impacting productivity and causing late deliveries. Culture erosion becomes more obvious when Technical Debt is not addressed in a timely manner. It manifests in multiple ways that include inability to staff a legacy project (software engineers choose best available assignments that leverage latest industry programming language(s), technologies, tools, and processes). Senior engineers, with vast domain knowledge, choose to leave the workforce leading to inability to adequately staff these outdated long-running programs.

Technical Debt extends from research and development projects, where annual decisions to continue or delay funding of multi-year software projects are made, to developmental and production programs, where programs determine whether to reuse existing software products or undergo new development. Technical Debt also spreads into decision-making associated with when to replace or upgrade developer tools and build environments, which are known to rapidly change.

The Boeing Enterprise Software Engineering organization has identified recurring Technical Debt examples and is now taking remedial steps to manage and eliminate them where possible. Two common examples are:

\begin{enumerate} 
    \item \textbf{Reified Prototyping}: Occurs when software solutions are built to support initial capability demonstrations and are then evolved into production solutions. The issue is that these preliminary prototypes were not designed and built to the same level of process rigor as required for safety-critical production programs, which results in high-cost re-engineering to meet the latter’s objectives.
    \item \textbf{Bespoke Designs \& Tools}: These software systems typically lack modularity, are not built with light-weight component interfaces and are tightly coupled with low cohesion. Such attributes lead to resistance to upgrade, where the designs, tools and build environments themselves become increasingly difficult to maintain.
\end{enumerate}

Strategies undertaken to combat these Technical Debt examples are often applied at the system level and address long-term business benefits. For example, bespoke designs are being replaced by feature-based product line engineering where software solutions in the form of products and capabilities are leveraged across multiple programs. Furthermore, open source software and modular tools are being composed and applied to enable frequent and incremental upgrade. As a broader measure, the Boeing Enterprise Software Engineering organization has established a Technical Debt Relief Bureau (TDRB) that is focused on raising Technical Debt awareness across the organization. The primary TDRB responsibility is to capture a consistent and continuously improving Technical Debt management methodology by:
\begin{itemize}
    \item Documenting Technical Debt items and root causes using established, uniform and consistent methods (as well as simple tools like spreadsheets and Markdown);
    \item Establishing a business model for Technical Debt mitigation;
    \item Performing holistic examination of the business model and key business goals relative to all software engineering decisions (e.g., shaping business objectives based on all potential future customers and product roadmaps while making full life cycle tradeoffs versus focusing on securing near-term business);
    \item Recommending corrective / mitigating actions;
    \item Applying new efforts and principles to prevent Technical Debt, e.g., common and modular development environments that enable incremental upgrade.
\end{itemize}

The long-term TDRB objective is inculcation of Technical Debt management via informed decision-making that naturally occurs and evolves across the Enterprise. The leveraging of open source tooling and consistent software architecture capture, e.g., using Krutchen's 4+1 Views \cite{Kruchten4plus1}, across the organization enables streamlined Technical Debt management and provides opportunities to evaluate and instantiate future Technical Debt management research advancements.

Known gaps and areas for further research:
\begin{itemize}
    \item Industry needs automated Technical Debt management tooling that provides broad stakeholder insight into its root cause and status. Required capabilities include: 1) configurable Technical Debt cost modeling that forecasts product-specific periodic (e.g., annual) maintenance funding levels required to adequately manage Technical Debt, 2) system-level  Technical Debt tracking and trending where the configurable cost model  is utilized to prioritize Technical Debt issues, e.g., filtered by incurring cost impact or by remedial cost and 3) relative to software product lines, tooling that can extract underlying Technical Debt root cause from unstructured Technical Debt, e.g., program tracked specific items, all of which trace back to a single root cause.

Common and consistent Technical Debt management tooling applied across all products will lead to the ability to make informed funding decisions across the organization. They will also  enable defense contractors to convey Technical Debt costs and share Technical Debt decision-making with the customer community; additionally, they should also provide rapid feedback on individual decision tradeoff choice, driven by organization-specific configuration file(s) and cost models.
    \item Industry must better exploit the relationship of product line engineering and Technical Debt. Large-scale software heavily relies on feature-based software engineering. Open source tooling exists for managing large-scale feature models; however, whether these tools adequately support Technical Debt tradeoff analysis remains as an unexplored topic.  Product line engineering requires frequent testing across all supported configurations. Tooling that can intelligently identify and execute a subset of tests based upon specific problem reports and code updates leads to earlier code error detection that can remedy multiple types of debt.
    \item Research aimed at increasing Technical Debt awareness and perhaps establishing a software community catalog of symptoms, causes and examples, searchable by software domain and/or life-cycle stage, could benefit practitioners by providing common understanding and insights. Existing research in these topics has not yet reached practical industry application. 
\end{itemize}

\subsection{Technical Debt and Legacy Systems Portfolio Management}

Legacy system portfolio management is a concern of not only large industry enterprises, but also governments across the globe. Technical Debt rapidly accumulates due to challenges tied to managing software systems through multiple decades of technology upgrades, handovers, and iterative functionality evolution. We use the US Department of Defense (DoD) as an example where many of  the challenges are representative of larger issues relevant to government software systems around the globe:
\begin{itemize}
    \item The agency is focused on its mission (whether civil or military) and tends to view its software assets as a means to accomplishing that mission, not end products; 
    \item Investments in such systems are made to accomplish public interest goals, not to create sources of revenue to pay for their upkeep or to contribute to the bottom line; 
    \item Systems tend to be long-lived and have evolved over the course of their lifetime, potentially by a series of different contractors; 
    \item Systems are potential targets of attack to a variety of adversaries, motivated by a spectrum of different reasons (e.g., to harm essential government functions, to uncover sensitive information or to harm national prestige). 
\end{itemize}

An increasingly important DoD cost driver is the effort expended to develop, acquire, and sustain software-intensive systems. Most DoD systems are in operation for long periods of time, and they continuously evolve to support incorporation of new functionality and to address maintenance issues. Even in mature systems, a continual demand for system adaptation is likely due to changing mission profiles, the need to incorporate more effective or efficient technologies into the system, or the desire to repair newly discovered software vulnerabilities. For all of these scenarios, software tends to be the logical and cost-effective way to make the upgrade, meaning that in effect, software continually evolves and is never “done.”

At a time when the scope of responsibility for many government agencies seems to be steadily increasing, the allocation of public funds between new systems / new capabilities and maintaining existing systems is increasingly fraught. However, we know that Technical Debt in many systems continues to accumulate even if the code is not being actively changed, e.g., because the underlying technologies become obsolete over time and the need  for periodic investments driven by modernization and associated software updates. These issues pose a continuous funding need to address Technical Debt that can be difficult to identify, given competing priorities.

Consequently, dealing with Technical Debt in software is an unavoidable phenomenon for the DoD. Similar to other organizations, for systems on which Technical Debt has been allowed to accumulate as the system evolves, dealing with a stream of continual changes becomes increasingly less cost effective as more and more effort is required to simply understand and modify the system, much less implement new capabilities. This scenario can result in cost escalation and schedule slippages and/or a diminished ability to field new capabilities. However, because these systems provide essential government and security services, the government typically has much less flexibility than a commercial entity regarding the decision to retire such software; it is not possible to simply stop maintaining a system because the cost of upkeep is no longer practical. 

Given that Technical Debt identification is often context-specific, individual teams have had a significant amount of leeway in defining their own practices around Technical Debt. As a result, there is a wide variety of different practices in place across the Department. However, similarly to the Boeing example above, there has recently been a renewed interest in raising awareness across the entire organization, and in supporting enterprise-level efforts to improve practices across the board. The 2019 report of a key advisory group, the Defense Innovation Board, emphasized the need to periodically invest in paying down Technical Debt~\cite{dib2019}. This led to a policy change requiring programs acquiring software for the Department to actively manage Technical Debt~\cite{dodi2022}.

The new policy places renewed emphasis on the need for software programs to demonstrate value by presenting periodic and frequent evidence about the impact on the mission from the working software they have produced. This approach will help make visible the instances where software efforts are hampered by large amounts of Technical Debt and unable to deploy new capabilities on meaningful timeframes. The Department of Defense’s 2021 Science and Technology strategy for software likewise re-emphasized the need to focus on Technical Debt reduction, linking it to the speed with which updates can be made to the software and hence to improving cybersecurity. This strategy focuses on research into Technical Debt approaches as a contributor to improving mission performance~\cite{dod-strategy}.

Government-driven research, such as the Technical Debt study required by the fiscal year 2022 US defense authorization act \cite{NDAASec835}, ensure that, while the teams acquiring individual systems make their own choices about how to manage Technical Debt in their unique situation, the Department can consider tools and practices that would improve performance across all teams. One issue that needs to be addressed, according to the terms of the study, is not just what measures to use to gain insight into Technical Debt, but whether and how to report a version of those measures to indicate progress to stakeholders and avoid surprises due to missed commitments or slow deployments of functionality. Other focus areas are likely to include how to propagate best practices across teams for reserving some amount of effort in each iteration to keep Technical Debt issues under control, and avoid the temptation to focus entirely on new capabilities.

Known gaps and areas for additional research include:
\begin{itemize}
    \item Ability to combine meta-data about the team with analysis of code base artifacts, to predict the levels of periodic maintenance that would be required to properly manage Technical Debt in a code base. Meta-data could describe aspects of the software effort such as the domain (e.g., business systems, logistics, data processing, embedded Cyber-Physical System software); length of time the software has been being evolved; quality metrics; etc. 
    \item Large enterprises need to manage Technical Debt across different teams and organizations and life cycle stages. Techniques that enable successful hand over of Technical Debt as teams and systems evolve can help improve open discussions around what Technical Debt stays in systems which needs to be resolved sooner than later. 
    \item Stakeholders who oversee a portfolio of different projects, need to understand how to allocate resources across those projects based on criteria that should include Technical Debt. For example, how can decision makers objectively compare different projects to understand which systems should be retired or replaced, if resources are available, and which have sufficiently managed their Technical Debt such that they can continue to add functionality?    
\end{itemize}

\subsection{Technical Debt, Speed and Quality For Organizations Developing Custom Software}

Like many small and medium enterprises that develop unique customer software solutions, QAware\footnote{QAware is a Germany-based  software manufacturer and consultancy developing custom software solutions} is challenged by intensified competition, driven not just by the sheer number of service providers but also by firms leveraging near and far-shore outsourcing to offer cost-competitive propositions. Given this context, consistently delivering superior quality and achieving high productivity are key aspects to remain relevant and ensure long-term clientele partnerships.

One strategy to achieve both high quality and high productivity is by establishing practices and principles for managing Technical Debt that ensure that quality is ingrained into software engineering and development practices as well as into project management. 

These practices and principles need to be flexible so that they can be adapted to different customer organizations and cultures. In addition, they need to take into account that, in agile settings, the responsibility for prioritizing the resolution of Technical Debt falls to the product owner, who is usually a customer representative. This allocation of responsibility implies that principles and practices cannot be dictated to product owners but rather need to be addressed to the product team, alongside techniques for making the prioritization arguments to the product owner.

The practices and principles also need to be comprehensive in order to cover Technical Debt and its interest as well as other debts and consequences: to provide prioritization arguments to the product owner, it helps to make consequences and interest tangible, such as implementation delays, risk, or product (in)stability.  Furthermore, being able to add value to a product over a long period of time (as opposed to a short-term project) requires addressing not only cost of change but also, for example, cost of operations or cost on the users’ side (such as user experience issues or costs caused because users need longer to complete workflows). 

Some examples of the principles and practices that QAware has established are: software craftsmanship and clean code principles through internal education and teaching measures that are mandatory for all developers; a quality contract as well as Technical Debt grooming and prioritization principles for all projects; and regular quality and innovation review at the organizational-level. 

Let us consider in more detail one of the aforementioned principles, namely the project-level \textit{quality contract} that includes continuous quality measurement based on available software quality tools, including, for example, SonarQube\footnote{https://www.sonarsource.com/} and OWASP\footnote{https://owasp.org/} . The contract enforces a zero-violation policy: any violations detected by these tools must be adjudicated and resolved. This zero-violation policy is critical to contain and prevent Technical Debt expansion due to quality erosion.  At the same time, it is often hard to achieve, as it may necessitate obtaining customer consent. 

One may argue that a zero-violation policy already detects and manages Technical Debt. However, existing tools focus on detecting low-level (software implementation) issues. This is valuable but, in QAware’s view,  only sets the basis for managing “relevant” Technical Debt and for achieving high quality. Figuratively speaking, the metaphor of software craftsmanship is often used to describe professional software development that aspires to achieve high software quality. Under this analogy, most tools that attempt to detect Technical Debt rather measure the amount of sawdust in the craftsman’s workshop. Thus, a zero-violation scenario is the equivalent of ensuring that the workshop floor is incrementally swept and remains clean of sawdust. While this is considered to be a prerequisite, true craftsmanship goes beyond the mere absence of such flaws.

Technical Debt management starts from there --  a solid, clean basis -- by: continually maintaining a refactoring backlog through manual collection of Technical Debt items from the team; manual estimation of their principal, and manual prioritization; and assurance that refactoring happens - be it through the Scout principle~\footnote{``Leave things better than you found them.''} or through explicit prioritization and planning. 

A particular challenge is to remain diligent on different causes for cross-cutting, “painful” Technical Debt. Examples are domain-level causes (e.g., misunderstood or changed user workflows) and systems architecture (i.e., on system landscape level, such as interfaces between different systems or products). These are difficult to detect and to address, since they require comprehensive knowledge and involve multiple stakeholders. However, when present, they can significantly reduce development speed and product value.

Current trends in custom software development increase challenges for detecting Technical Debt. With the growing importance of DevOps and cloud applications, software development teams must build expertise in deployment and monitoring processes as well as platform engineering. This situation implies that code bases contain new artifacts - along with new programming languages such as Terraform - to support these tasks. At the same time, legacy programming languages grow in importance: Many legacy applications are currently either migrated to cloud applications, or companies seek external support for maintenance of legacy systems. One driver for this trend may be retirement of many expert developers. Furthermore, because the trend is to develop landscapes of (micro-) services instead of monoliths, application code is becoming distributed across many repositories, while measurement tools tend to focus on single repositories. Last but not least, artificial intelligence (AI) is increasing in importance; this means that we need to learn how to express and measure Technical Debt of AI-enabled systems (i.e. systems that contain at least one AI component inside). 

Known gaps and areas for additional research include:
\begin{itemize}
    \item Better support for detection of debt with high impact, e.g., early life cycle architecture Technical Debt: tool support is currently typically restricted to downstream, low-level issues (code, low-level design). While this is highly valuable, it falls short of detecting high-level design and architecture issues, which are often cross-cutting and occur early in the software life cycle, and therefore are typically the most impactful and “painful” ones
    \item Support Technical Debt detection for emerging programming languages (e.g., Rust, Julia, Terraform) and older programming languages in legacy code (e.g.,  Ada, Cobol).
    \item Support for detection and management of Technical Debt symptoms and tracing them to the respective Technical Debt item(s).
    \item Support for expressing and communicating interest and other consequences and costs.
    \item Understanding and representing applications as sets of repositories; that is, sets of microservices that include application, platform, and operations code.
\end{itemize}

\subsection{Technical Debt Ownership by Business and Architecture in Large Organizations}
CGI\footnote{CGI is a large global IT services and business consultancy organization} sees many of its client organizations struggling with Technical Debt. 79\% of CIOs interviewed worldwide by CGI indicate that their ability to change is slowed significantly by technology and agility constraints. CGI’s consultants see that these constraints are often related to Technical Debt. High pressure to produce new features causes business stakeholders to consistently under-prioritize the work needed to keep IT landscapes healthy and up to date. An informal analysis based on anecdotal evidence yielded the following potential causes:
\begin{itemize}
    \item Short term focus caused by quarterly result pressure and associated Key Performance Indicators;
    \item Overinflated stakeholder expectations caused by short ‘pilot’ projects that produce quick business results, but not at a sustainable rate;
    \item Misapplication of agile practices without understanding their intended context or limitations; examples are “Weighted Shortest Job First” (a prioritization formula that works well for business features, but is inadequate for  Technical Debt repair and other backlog items that add only indirect business value, so-called `enablers') and “You Ain’t Gonna Need It” - YAGNI (a principle to prevent unnecessary future-proofing of a product);
    \item Splitting budget between business and IT (business pays for new features, IT pays for ‘maintenance’ including Technical Debt management).
\end{itemize}

Business stakeholders are often surprised by the size of the Technical Debt, and sometimes blame IT for not keeping it under control. Part of the problem is in the name: because it’s called \emph{technical} debt, business stakeholders are not inclined to take problem ownership; they rather feel like victims of a technology problem.

One thing that becomes clear from the aforementioned causes, is that \emph{keeping Technical Debt under control cannot be achieved by technical departments only}. The decision tradeoffs that involve Technical Debt must be made by collaborative stakeholders from both business and IT. This makes Technical Debt management part of business-IT alignment, which in turn is an architecture responsibility. This is why CGI has added Technical Debt management practices to its architecture approach, called Risk and Cost Driven Architecture \cite{Poort2012}. The practices are focused on fostering collaboration between business owners and development teams by:
\begin{itemize}
    \item Translating Technical Debt related concerns into business terms like risk, cost of delay and opportunity cost;
    \item Making the tradeoffs in decisions involving Technical Debt transparent (such as when to eliminate the debt by refactoring or upgrading), e.g. in business cases;
    \item Visualizing the dependencies between Technical Debt, business features, product quality and external pressures in ways that are easily digestible by business stakeholders \cite{Poort2016}.
\end{itemize}

Reports from CGI architects show that such practices help to involve business stakeholders in the decision-making process; particularly, once they realize that Technical Debt and enablers are an integral part of their products, they start to feel ownership of Technical Debt. This in turn helps prioritize Technical Debt resolution and management at the business level. 

Important areas of further research include:
\begin{itemize}
\item The relationship between the technical and business/economic aspects of software evolution needs attention. We need a deeper understanding of the business impact of the timing of Technical Debt remediation, taking into account opportunity cost, team velocity and sustainable rates of software development. It would help to have funding and budgeting models that clarify the real tradeoffs in the business decisions surrounding Technical Debt \item While industry focuses on architecture in managing Technical Debt, the practices developed in industry need validation: evidence-based practices will help in the (sometimes painful) process of getting business stakeholders on board. 
\end{itemize}

\subsection{Themes from Industry Perspectives}
The four industry perspectives shared in this section come from very different domains and organizations, but reveal a number of common themes:

\begin{itemize}
    \item \textit{Value generation \& ROI.} The bottom line of Technical Debt management is return on investment (ROI) for the business and value, as difficult as it is to quantify financial bottom line reliably.
    \item \textit {Architecture in addition to code.} All perspectives emphasize the difficulties around managing architecture-level Technical Debt where the tradeoffs are more implicit and more complex. 
    \item \textit{More capable tools.} All perspectives suggest that in the absence of sophisticated tools, they have repurposed code quality tools to help manage Technical Debt, but are aware of their gaps and desire better tools.
    \item \textit{Common principles and practices.} Organization specific  principles and established practices for managing Technical Debt exist, but are not codified.
    \item \textit {Continuous Practice.} All perspectives demonstrate that detecting and resolving Technical Debt is not a one time activity and involves input from multiple stakeholders. 
\end{itemize}

Next, we will look at the state of research. The state of both industry and research, and particularly their shortcomings and relevance to each other feed into the vision in Section \ref{sec:vision} as that will require the contribution of both researchers and practitioners.

\section{Reflections from more than a decade of Technical Debt research}
\label{sec:research}

One of the indicators of the progress and maturity in a scientific field is the number of peer-reviewed research publications investigating it. In the case of Technical Debt, since the very first ``Managing Technical Debt" workshop in 2010 \cite{Brown2010}, hundreds of related publications in scientific conferences and journals have been released. 
The latest tertiary study in this field \cite{Junior2022} summarizes the results of 19 secondary studies (systematic literature reviews or mapping studies) that in turn collectively summarize 532 unique primary studies.

In this section we reflect on the progress research has made on developing tools, techniques, and practices for managing Technical Debt. It is not our goal to conduct another systematic literature study. Instead, we present an informal meta-analysis of the literature from both the secondary and the tertiary studies to map the existing evidence, and summarize the main research contributions and trends. Specifically, we use as input all secondary and tertiary studies (see Table \ref{tab1}) on the field of Technical Debt contained within CORE A scientific journals, as those set a certain quality expectation. 

Our meta-analysis reviews the current state of research by summarizing focus areas that received most attention, as well as related fields. Subsequently, we review data sets, and tools followed by implications of current research.  

\subsection{Research Focus Areas in Technical Debt}

A majority of the research studies have been exploratory, attempting to understand the different facets of Technical Debt and its management, define the related concepts, recognize current industry practices and the needs of software practitioners, and propose potential technological solutions for managing Technical Debt. In this respect, significant progress has been made on both defining the problem space and in exploring the solution space. 

Research focus areas that received most attention include definitions, conceptual models, classifications of Technical Debt, and the Technical Debt management process.

\begin{table}[htbp]
\caption{Overview of SLRs and SMS related to Technical Debt (journal publications with at least CORE A)}
\begin{tabular}
{|p{7em}|p{9em}|p{1em}|p{1em}|p{1em}|}
\hline
\multicolumn{1}{|c|}{\textbf{Authors}} & \multicolumn{1}{|c|}{\textbf{Title}} & \multicolumn{1}{|c|}{\textbf{Journal}} & \multicolumn{1}{|c|}{\textbf{Year}} & \multicolumn{1}{|c|}{\textbf{\#Studies}} \\
\hline
HJ. Junior, GH. Travassos &
Consolidating a common perspective on Technical Debt and its Management through a Tertiary Study & IST & 2022 & 19 (secondary) \\
\hline
A. Melo, R. Fagundes, V. Lenarduzzi, W. Barbosa Santos &
Identification and measurement of Requirements Technical Debt in software development: A systematic literature review & JSS & 2022 & 66 \\
\hline
V. Lenarduzzi, T. Besker, D. Taibi, A. Martini, F. Arcelli Fontana &
A systematic literature review on Technical Debt prioritization: Strategies, processes, factors, and tools & JSS & 2021 & 38 \\
\hline
N. Rios, M.G. de Mendonça, N. Rodrigo, O. Spínola &
A tertiary study on Technical Debt: Types, management strategies, research trends, and base information for practitioners & IST & 2018 & 13 (secondary) \\
\hline
T. Besker, A. Martini, J. Bosch &
Managing architectural Technical Debt: A unified model and systematic literature review & JSS & 2018 & 42 \\
\hline
W. N. Behutiye, P. Rodriguez, M. Oivo, A. Tosun &
Analyzing the concept of Technical Debt in the context of agile software development: A systematic literature review & IST & 2017 & 38 \\
\hline
C. Fernández-Sánchez, J. Garbajosa, A. Yagüe, J. Perez &
Identification and analysis of the elements required to manage Technical Debt by means of a systematic mapping study & JSS & 2017 & 63 \\
\hline
N. S.R. Alves, T. S. Mendes, M. G. de Mendonça, R. O. Spínola, F. Shull, C. Seaman &
Identification and management of Technical Debt: A systematic mapping study & IST & 2016 & 100 \\
\hline
Z. Li, P. Avgeriou, P. Liang &
A systematic mapping study on Technical Debt and its management & JSS & 2015 & 94 \\
\hline
A. Ampatzoglou, A. Ampatzoglou, A. Chatzigeorgiou and P. Avgeriou &
The financial aspect of managing Technical Debt: A systematic literature review & IST & 2015 & 69 \\
\hline
\end{tabular}

\label{tab1}
\end{table}

\textbf{Definitions}. Early studies (2010-2015) attempted to provide a clear definition of Technical Debt according to existing literature, practitioners’ insights or (empirical) theory building. Up to that time, the term was rather ambiguous, with different authors interpreting it flexibly and placing emphasis on diverse causes and impacts of poor software quality. It is noteworthy that 107 definitions of Technical Debt were retrieved from the literature in 2015 \cite{Poliakov2015}. This quest to define Technical Debt significantly slowed down when the 2016 Dagstuhl seminar produced a definition that became well accepted in scientific literature~\cite{DagstuhlAvgeriou2016}. 
The Dagstuhl definition is sometimes debated regarding the semantics of the concepts (see Section~\ref{sec:whatis}), but the definition itself has stood the test of time in the research community and is broadly referenced. 
In contrast, software practitioners still embrace a rather loose, and sometimes ambiguous Technical Debt definition, also often referring to Cunnigham's original definition; this is also the case in all four organizations that are discussed in Section \ref{sec:industry}.  

\textbf{Conceptual Models}. Numerous studies have attempted to provide a conceptual model of the term that explains the main notions, such as:
\begin{itemize}
    \item Concepts from the debt metaphor (e.g., principal, interest, interest probability, value, bankruptcy, investment)
    \item Causes (e.g., limited budget, lack of system knowledge, business pressure)
    \item Benefits (e.g., gain business opportunities, accelerate software release)
    \item Consequences (e.g., risks, reduced development velocity, increased cost of change, market loss) 
    \item Artifacts of the software development life cycle affected by debt (e.g., architecture, design, code, test cases, documentation)
    \item Symptoms (e.g., code smells or code complexity)
    \item How it occurs (e.g., intentional vs. inadvertent, prudent vs. reckless)
\end{itemize}

While the 2016 Dagstuhl seminar also produced a conceptual model with the main relevant notions~\cite{DagstuhlAvgeriou2016}, we have seen several such models in recent secondary studies that either extend existing models or specialize them for specific debt types. Notably, several studies point out the limitations of the metaphor and particularly how it differs from financial debt (e.g., Technical Debt interest is not always paid); this observation has raised community awareness that the debt analogy is a useful means of communication but should not be over-extended to all possible related financial concepts and needs its own software economics model.   

\textbf{Classifications of Debt}. Technical Debt has been classified into different \textit{types}. While these classifications are not identical or aligned, some types consistently repeat across studies, e.g., architecture, design, code, test and documentation debt. Some types are often misunderstood. For example, “defect debt” does not refer to the existence of actual defects but to knowing of defects and deferring their fixing, as explained in Section \ref{sec:whatis}. 
Similarly, “requirements debt” does not refer to unimplemented features but unclear, incomplete or inconsistent requirements specifications that adversely affect software maintenance. The boundaries of several types are often unclear; for example, “design debt” is often not differentiated from “code debt”, simply because its identification is based on the analysis of source code.

\textbf{Process of Technical Debt Management}. Unlike outcomes of research on conceptual models or types of Technical Debt, there is wide consensus on the various activities constituting Technical Debt management, with the list from Li et al.~\cite{Li2015}, being very broadly used: repayment, identification, measurement, monitoring, prioritization, communication, prevention, representation/documentation. Most of the existing approaches on Technical Debt management have focused on identifying and measuring debt, and to a lesser extent on representing, monitoring, and repaying it. Measurement predominantly concerns the principal, while Technical Debt interest is far more challenging to measure \cite{Martini2017}. Among these approaches, many propose tools (see Table \ref{tab:tools}) while others suggest practices (e.g., dedicated code reviews, or including Technical Debt items in Agile team backlogs). There are several approaches that utilize economics theory (e.g., cost benefit analysis or real options) to exploit the metaphor of debt; although the adoption of such approaches is rather limited in practice, they have broadly established the idea that Technical Debt is not only a liability but also an investment that can be of immense value to an organization.

Despite the sound understanding of the Technical Debt management process, current research is characterized by an almost complete absence of work in human and social aspects. This is a striking paradox, as Technical Debt is, by and large, the result of human actions and decisions, both deliberately and inadvertently. Furthermore, Technical Debt needs to be understood by humans and needs to be discussed and acted upon by different stakeholders. This has been validated by both the exploratory research studies regarding the symptoms, causes and consequences of Technical Debt management, as well as the industry challenges identified in Section \ref{sec:industry}.

\subsection{Related fields}

Despite being a relatively new subfield of software engineering, Technical Debt research has been published in various journals and conferences.
To obtain a representative overview of the Technical Debt landscape, we analyzed 193 primary studies from the pool of papers in \cite{Junior2022}, appearing in at least two secondary studies. This inclusion criterion was applied to ensure both the relevance of each primary article to the field of Technical Debt and the research rigor of the work. The majority of research papers (24\% of the 193 primary studies) have been published in the International Conference on Technical Debt (TechDebt) which started in 2018 and its predecessor, the International Workshop on Managing Technical Debt (MTD), first organized in 2010. In terms of publication type (journals/conferences/books), 73\% of the papers on Technical Debt appear in conferences. 

Over the years, Technical Debt articles started appearing in venues associated with neighboring fields, such as the Mining Software Repositories conference (MSR) and the International Conference on Software Maintenance and Evolution (ICSME). In addition to MSR and ICSME, papers related to Technical Debt now frequently appear in industry and research tracks of top software engineering conferences such as the International Conference on Software Engineering (ICSE) and Foundations on Software Engineering (FSE) as well as topic-specific conferences such as International Conference on Software Architecture (ICSA), International Conference on Program Comprehension (ICPC), Empirical Software Engineering and Management (ESEM) and International Conference on Software Analysis, Evolution and Reengineering (SANER). 

Such widespread research publication is reasonable, since the tools/methods being employed for identifying, measuring, and repaying debt often fall under the scope of software analysis and reverse engineering for improving maintainability. It is also an indication that \textit{Technical Debt management has become mainstream} within the software engineering research community. From the remaining venues that publish Technical Debt research, it becomes evident that agile, architecture, and maintenance communities are increasingly concerned with the long-term impacts  of shortcuts taken on software maintenance and evolution. As a final note, we remind that the primary focus of our analysis was peer-reviewed publications; with that said, Technical Debt has become a regular topic in industry conferences and gray literature.  

\subsection{Tools}

Many tools measure properties of software and related artifacts, beginning with early measurement activities characterized by work on project management \cite{Boehm1981}. While a significant amount of research focused on tools, it was mostly on applying commercially available tools and analyzing the outcomes for indications of debt. There has been a lack of agreement about what distinguishes a `Technical Debt tool' from one which is a static code analyzer. These static analysis tools are often erroneously assumed to be a complete solution for Technical Debt management. 
 Tools which focus specifically on Technical Debt should have at least the following properties:
\begin{itemize}
    \item Uses the Technical Debt metaphor and identify a Technical Debt Index which defines principal (the cost to refactor or repay the initial debt) and/or interest (additional costs incurred by the project in the presence of Technical Debt). 
    \item Enables tradeoff assessment for different decisions in software development.
    \item Focuses on code, design, tests, or architecture.
    \item Is commercially relevant. 

\end{itemize}
This is not exhaustive as Technical Debt tools should be able to perform a variety of tasks, including finding, measuring, fixing, or prioritizing debt.

This list of properties excludes the plethora of studies that examine defect prediction, code smell analysis, or refactoring support. While clearly relevant to identification and  management, these studies do not explicitly focus on Technical Debt. Table~\ref{tab:tools} lists tools which have features which assist Technical Debt management based upon these criteria. Note that the table is not exhaustive and does not reflect our endorsement of any particular tool.

Tools primarily automate the detection and measurement of indicators of Technical Debt. Measurement usually takes the form of either static analysis rules or dependency measures of source code connections. Less frequently, measurement uses software process measures (such as commits over time) based on activity from social coding sites such as Jira or GitHub. For example, many Java project tools are based on variants of the original FindBugs Java code quality ruleset\footnote{\url{https://findbugs.sourceforge.net/bugDescriptions.html}}; for Python, rules might reference the PEP8 formatting standard or other such `linter' rules. 

Some of the existing tools do an excellent job in identifying rule violations and non-compliance to principles at the code level, thereby promoting a culture of 'clean code' within development teams.  
However, a significant number of research has  identified the needs of software practitioners in terms of usable tools that effectively manage Technical Debt and those needs are largely unmet. Very few tools focus on Technical Debt in architecture and design (e.g., DV8\footnote{\url{https://archdia.com}} and Arcan\footnote{\url{https://essere.disco.unimib.it/wiki/arcan/}}), which has a much larger impact on the software project and team as compared to downstream code; for example most tools give little insights other than how problematic big files are. Furthermore, current tools do not offer enough automation, and treat Technical Debt management activities as one-off exercises instead of continuous. Finally, current tools do not integrate well with other open source tooling and workflows used in practice. 

In many cases, tools are an entry point for a bigger consulting and change management project. Tools such as CAST or SonarQube are complex enough to require tool experts to properly configure them (e.g., by choosing what source code to include in the analysis, which plugins to use, what rulesets to use). Most such tools permit ruleset customization, and the SARIF standard~\footnote{\url{https://www.oasis-open.org/committees/tc_home.php?wg_abbrev=sarif}} aims to make such rulesets interchangeable. 

However, rules are not all equal, and tools rarely agree on the amount of identified debt \cite{Amanatidis2020,Lefever2021}, since the amount of customization changes---both in which rules to enable, and how important a violation is. Thus, a Technical Debt Index as a construct is very difficult to standardize: what constitutes critical Technical Debt, mostly based on code quality and related metric threshold violations, for one tool may not even be highlighted by another. False positives are also not uncommon, and yet not easily predicted. From a research perspective, however, some form of agreement is critical if in-depth analysis of these tools---for example, whether they agree on total debt for a given software system---is to be useful.

\begin{table*}[]
     \caption[short]{ A Summary of Popular Tools Used to Manage Technical Debt, extending the list from~\cite{Avgeriou2021AnOverviewComparison}   }
    \begin{tabular}{@{}llll@{}}
    \toprule
    Tool (date created)         & Focus                          & Languages                      & Technical Debt Index Definition                                              \\ \midrule
    CAST (1998)                 & Code, design, architecture     &          Most                  & Violations * rule criticality * effort                      \\
    SonarGraph (2006)           & Design, Arch                   & Java, Kotlin, Python, C\#.     & Structural debt index * minutes to fix                      \\
    NDepend (2007)              & Code, design, arch             & .Net frameworks                & Violations * fix effort                                     \\
    SonarQube (2007)            & Code                           & Most, with plugins             & Cost to develop 1 LOCe * Number of lines of code.           \\
    SQUORE (2010)               & Design, code                   & C++, Java, others with plugins & N/A                                                         \\
    DV8 (2019)                  & Architecture                   &                                & Penalties: additional bugs and/or changes in lines of code. \\
    Silverthread CodeMRI (2013) & Design                         & C++, Java, Fortran, Cobol      & N/A                                                         \\
    Symfony Insight (2019)      & Code and dependencies & PHP                            & Number of issues * time needed to remove the issue   \\ 
    CodeScene   (2017)                & Process, code, design          & Many                           & Development activity + rule violations                      \\
    Arcan         (2015)          & Architecture                & Java, C\#                       &     Severity and extent of architecture smells               \\
    Designite         (2016)          & Code and design                & Java, C\#                       &     Design rules violated                                                        \\ \bottomrule
                                                     
    \end{tabular}
    \label{tab:tools}
    \end{table*}

\subsection{Data Sources and Datasets} 
Studies to date, in particular those proposing Technical Debt management approaches, have used a variety of data sources, mining not only source code repositories and bug tracking systems, but also documentation, design and requirements specifications, test reports, backlogs, commit data, change/pull requests, etc. 
In addition to automated techniques, several studies use case study as a research strategy to close information gaps, which exist when only mining the aforementioned data sources; often practitioners are surveyed in such studies and their opinions are related to work artifacts. 

Most Technical Debt management approaches also use indicators or symptoms of debt as input, instead of Technical Debt items per se, since it is sometimes difficult or even impossible to identify said Technical Debt items. Thus, most Technical Debt management studies choose to rely on identifying indicators such as code smells, metrics (e.g., coupling, test coverage), dependencies, code complexity, etc. as a proxy for Technical Debt items. One exception to only considering indicators is Technical Debt management procedures focusing on \textit{self-admitted debt} \cite{Potdar2014}: problems, inefficiencies and pending improvements which are explicitly documented by the development team in source code comments, issue trackers, pull requests, commit messages, etc. and can be detected using various methods. 

One of the typical goals of most secondary and tertiary studies is to examine the coverage of different focus areas over time to identify which areas are under-studied. The studies agree that code has been extensively researched for identifying Technical Debt, that of design and architecture has received some attention, and to a lesser extent that of test and documentation \cite{Junior2022}, \cite{Rios2018}. 

When one looks at existing tools to perform Technical Debt management, the predominant focus is on code. This comes in stark contrast with the significance of the latter compared to architecture and design, which is considered the main pain point in Technical Debt management~\cite{Ernst2015MeasureItManage}. While several primary and secondary studies emphasize the need to move away from code to study Technical Debt in architecture\cite{Li2015}, \cite{Alves2016}, \cite{Behutiye2017}, \cite{Besker2018}, the trend shows the opposite: an even further increase of studies on code and Technical Debt. This is understandable: code is more concrete; source code is often the most reliably up-to-date artifact; and perhaps most importantly, it is far easier to collect data on. Design and architecture often involve incomplete or out-of-date documentation, poorly structured design rules, and consequently offer very scarce data to mine Technical Debt items (and subsequently use in relevant empirical studies).

 Despite challenges, there are some prominent examples of data used in research that have served as baselines or shared community infrastructure as representative datasets. As illustrative examples, we list some datasets that specifically identify Technical Debt Index \cite{verdecchia2022empirical} or Technical Debt discussions as the unit of analysis  in Table \ref{tab:datasets}. Overall, existing Technical Debt datasets found in research reveal the following characteristics: 
\begin{itemize}
    
    \item All datasets focus on open source, often open source ecosystems (e.g., Apache). In some cases, a paper with a tool will report on proprietary data but the data are not broadly shared~\cite{Tornhill2022CodeRed}.
    \item The datasets are nearly all Java-based. While Java is still a dominant language, particularly in business software, we have very little insight into tool performance on older languages (e.g. ADA) or newer languages (e.g. Python, Go). 
    \item Most papers build their own datasets, making it very difficult to compare research results. Analyzing open source projects with tools such as SonarQube or Codescene is possible with GitHub integration. With tools such as Sonarlizer~\cite{Pina2022SonarlizerXplorer} and repository API access, building a dataset is easier in many ways than downloading and reusing a third-party dataset.
    \item Datasets are not validated by humans involved in the projects and so the connection to real, impact-causing Technical Debt is hard to quantify. 
    \item The underlying approaches to build a Technical Debt dataset are highly related: rule-based code linters and software dependency networks based on different notions of dependency (such as package imports and method calls). This means that building datasets can be standardized and reused to facilitate researchers and also allow for evaluating their validity.
    \item Datasets based on research into code smells or refactoring detection are highly relevant, but not much used in Technical Debt studies. 
    \item There is a disconnect between what a tool might identify as a Technical Debt item or Technical Debt Index, and whether the developers and other stakeholders on the project would concur with what the tool finds. Some studies (e.g.,~\cite{Fang2022Cider}) report on human approval of the approach, but this is different than a tool which makes its way into the organization's toolchain (presumably adding great value).
\end{itemize}

\begin{table}[]
     \caption{Representative Technical Debt Datasets}
    \begin{tabular}{@{}lp{5cm}@{}} \toprule
    Dataset (with hyperlink)                                                     & Components                                        \\ \midrule 
    \href{https://zenodo.org/record/7757462}{PENTACET}~\cite{sridharan_murali_2023_7757462}     & Labeled SATD code comments from Java OSS projects \\
    \href{https://zenodo.org/record/4588039}{Lefever, tool comparisons}~\cite{Lefever2021}    & Consensus among tools                             \\
     \href{https://github.com/S2-group/ATDx_replication_package}{ATDx}~\cite{verdecchia2022empirical}   &   Architecture Technical Debt Index                                   \\
    \href{https://zenodo.org/record/3979784}{Amanatidis} ~\cite{Amanatidis2020}            & Agreement among tools on a Technical Debt Index    \\ \bottomrule                      
    \end{tabular}
    \label{tab:datasets}
\end{table}

\subsection{Themes From the Current State of Research}

The analysis of the scientific literature, the tools and the datasets shared in this section  reveal a number of common themes, similar to and overlapping with the industry themes in Section \ref{sec:industry}:

\begin{itemize}
    \item \textit{Code prioritized above architecture.} The proverbial lamp post research tends to focus on the easy part (code analysis) rather than the much more valuable but also challenging part (architecture) to analyze Technical Debt.
    \item \textit{Tools fall short.} Tools that are currently used in practice offer a starting point for discussion, instead of an in-depth analysis of important Technical Debt items. 
    \item \textit{Inadequate datasets.} We currently lack standardized datasets (that allow comparison between tools) and pertinent datasets (beyond Java projects and Open Source systems)
    \item \textit{Unclear scope.} There is still use of different Technical Debt types (e.g. defect debt, requirement debt) in an ambiguous way, hampering research progress. Also, other kinds of debt (e.g. security, social or AI debt) are used without being defined and without clarifying their relation to Technical Debt, and lead to fragmentation of research.
    \item \textit{Weak validation.} Most of the tools and the datasets are not validated with human subjects, casting doubt about their validity and their relevance to industry.
    \item \textit{Narrow research.} There isn't sufficient  emphasis on the social side of Technical Debt management, while research in a number of related fields is mostly isolated.
\end{itemize}

\section{Vision for Essential Technical Debt Management}
\label{sec:vision}

We have analyzed the scientific literature, tools for managing Technical Debt, research datasets, and most importantly the state of practice perspectives in four representative industrial organizations. These have guided us to derive common themes both from what the industry needs (end of Section \ref{sec:industry}) and what the research currently lacks (end of Section \ref{sec:research}). 
Based on these themes, we next present our vision for \textbf{how Technical Debt management should be practiced and researched} in five years from now. This vision consists of a number of points; Table \ref{tab:summary}, summarizes these points and how the themes from industry and research map to them. 

\visionhead{Technical Debt as Value Creation}
Software teams will learn how to effectively apply emerging Technical Debt management techniques, tools, and proven strategies to avoid debt unnecessarily creeping into systems.  
Having said that, inherent in the definition of any type of debt, and specific to Technical Debt as we have defined it, incurring debt can create \textit{value}. 
For this to occur, the practices and discussion around managing Technical Debt will shift to conscious value-add tradeoff management, and will be conducted collaboratively with business stakeholders. Development teams will be empowered to collaborate by tools that help them "translate" Technical Debt and its consequences into business terms such as ROI.

One of the largest promises of Technical Debt, perhaps not always well communicated, is that it can and should be positively applied, to capture value (analogous to a monetary loan for a useful purchase). It is imperative to raise awareness among software engineers, architects, and stakeholders of this shift in the perception of Technical Debt as “bad quality” to a “value generation and beneficial management” perspective. A shift to a value creation perspective in managing Technical Debt~\cite{Banker2021} will enable more powerful financial models to emerge from research and practice and assist ROI of taking on, keeping, and removing Technical Debt decisions. This positive attitude will help business stakeholders to take co-ownership of Technical Debt management.

\visionhead{Prioritize Architecture Above Code}
Technical Debt management will focus on early development decisions, especially architectural ones, that will be proactively analyzed and managed across the software life cycle. The priority will shift from low-level code issues to more expensive (difficult to change) decisions that are explicitly acknowledged as causes of Technical Debt items and prominently placed in the backlog. By architecture, we refer to multiple abstraction levels including system context and system landscape, as well as tradeoffs among technology and design choices.
Many organizations will have adopted the use of standardized architecture views, e.g., applying Krutchen's 4+1 Views \cite{Kruchten4plus1} and extending to other appropriate ones \cite{DSABookSEI}, leading to more concrete architecture artifacts where tooling can consistently identify and track Technical Debt items.

Development tools and teams alike will have incorporated holistic tradeoff analysis into all major and irreversible decision milestones encountered over the system life. Teams will routinely review, forecast, and identify areas where costly Technical Debt could occur, and act proactively. Approaches on mining self-admitted architecture Technical Debt from a variety of sources such as code comments, issue trackers, pull requests, and commit messages are complementary to source code analysis, and will enable Technical Debt detection in architecture that is not explicit in source code (e.g., poor design decisions or outdated third-party components). The reactive recognition and disposition of large Technical Debt items only after they become too costly to ignore will have become an unsustainable relic of the past. 

\begin{table*}[]
\caption{Mapping between industry needs (Section \ref{sec:industry}), research gaps (Section \ref{sec:research}), and vision for Technical Debt Management (Section \ref{sec:vision})}
\label{tab:summary}
\begin{tabular}{@{}p{4cm}p{6.5cm}p{6.5cm}@{}}
\toprule
Needs from Industry & Research shortcomings & Vision \\ \midrule
Value generation \& ROI &   & Technical Debt as value creation \\
Architecture in addition to code & Code prioritized above architecture & Prioritize architecture above code \\
More capable tools & Tools fall short & Technical Debt management using next-gen tools\\
Continuous practice &   & Technical Debt management reflects continuous practice \\
Common principles and practices &  Inadequate  datasets; weak validation; unclear scope & Technical Debt research is data-driven and relevant\\
 & Narrow research & Technical Debt research is socio-technical and multi-disciplinary \\ \bottomrule
\end{tabular}
\end{table*}
\visionhead{Technical Debt Management using Next-Generation Tools }
We currently experience rapid technology advancements associated with AI-based capabilities and tooling automation. These revolutionary changes have the potential to make incremental progress in how  Technical Debt management is orchestrated. Advanced tooling will assist with developing and analyzing code; developers will  avoid implementation mistakes or catch them much sooner. Consequently, effort spent on managing code Technical Debt will diminish, thereby providing opportunities to explore more targeted detection and reduction of more important, architecture Technical Debt. Tools which automate Technical Debt identification, measurement, documentation, prioritization, monitoring and repayment activities will be incorporated into developer routine workflows. 

Increased levels of tool sophistication and automation will enable common and consistent Technical Debt management across projects where organizations will use these data to make informed decisions on where to best apply limited maintenance budgets. Proactive Technical Debt management will enable continuous monitoring and scenario analyses that can convey implications of business in Technical Debt management decision-making.

Current tools which analyze source code files to detect a limited set of dependency-based architecture smells and other structures (e.g. DV8 and Arcan) will expand their capabilities to analyze a broad set of smells and architectural anti-patterns. They will seamlessly and timely track the evolution of Technical Debt over time and will enable monitoring of risky and costly architectural problems. 

AI-augmented and other tools, based on generative AI or other methods alike,  will provide features such as:  automatically detecting debt in code, comments, issues on the fly (i.e. as they are written) and documenting the Technical Debt item (even as simply as tagging an issue as Technical Debt); preventing debt by prompting the developer whether it is worth taking the shortcut or whether an alternative good solution should be considered; helping to prioritize repayment depending on the available resources and the ROI; (semi-) automatically repaying debt by following automated program repair, suggesting refactorings and implementing them. Such tools will be trained on appropriate code and will not create additional debt. 

\visionhead{Technical Debt Management Reflects Continuous Practice} 
Technical Debt management will be fully incorporated into a continuous software development life cycle. 
A shift to intentional, data-driven, and well documented Technical Debt will occur, resulting in concrete discussions around how to react when the debt stops being of value, what attributes, practices and tools are needed to monitor it, and in what intervals to revisit the decisions with ample and domain-specific data to support the Technical Debt management process. Such continuous practice will be supported by dashboards for a wide variety of stakeholders enabling improved insights into system development and evolution.

Resulting best practices from practitioners for this continuous management will be codified and applied broadly across industry, similar to other workflows in software engineering such as continuous integration/continuous deployment. Principles of Technical Debt management, such as translating technical results to business language, will be extended, and routinely exercised as a core part of disciplined software engineering further strengthening broadly accepted continuous management of Technical Debt in software organizations.  

\visionhead{Technical Debt Research is Data-driven and Relevant} 
Sophisticated tools and especially those supporting the aforementioned continuous practice will have produced rich industry datasets, expanding on the examples in Table \ref{tab:datasets}. New datasets will not only be readily available, but will also complement open source code with industrial systems, in a variety of languages and contexts. They will therefore represent the complexity of Technical Debt issues in practice, reflecting the progressive phases  that accumulated Technical Debt items undergo as systems evolve. 

In addition, datasets will be curated following empirical principles and practices while being validated with the help of practitioners in the field with longitudinal studies. All datasets will enable transparency and reproducibility by identifying which practices enabled their collection, as well as what practices or principles are represented in them at what point in time. Datasets will represent not only the most tangible outcome of the cross-pollination between industry and research, but also the value of incorporating Technical Debt management to software development and evolution from cradle to grave.

Going forward, Technical Debt management research will be cognizant to avoid `lamp post research' and scope work with awareness of industry challenges. The discussion on what is and what is not debt will have ceased.  The research community will turn its attention to solving practical problems negatively impacting industry. Studies classifying types of debt will simply reflect the artifact being studied in reference to Technical Debt in a given study. Both researchers and practitioners will systematically put into use the aforementioned litmus test (see Section \ref{sec:whatis}) on whether an item is classified as Technical Debt: does it incur interest when being changed? 


\visionhead{Technical Debt Research is Socio-Technical and Multi-disciplinary}
 
Research in Technical Debt will address the social aspects on equal grounds with the technical ones. The reasons that software practitioners incur Technical Debt will have become crystal clear, while supporting both individuals and teams in preventing, or eventually managing the accumulated debt from a socio-technical point of view will be the norm. The body of knowledge in software engineering research regarding human and social aspects will be well utilized in investigating how those aspects affect the way Technical Debt is incurred and eventually managed. New areas of research to understand how to reduce or eliminate systematic causes of Technical Debt will emerge. 

At the same time, a number of fields will be involved and contribute, each from its own perspective, to developing new theory and tools in Technical Debt management: Mining Software Repositories, Maintenance and Evolution, Empirical Software Engineering, Program Analysis, Automated Software Engineering, Software Architecture, Refactoring/Repair, Human Aspects of Software Engineering, Software Economics. Research with all those different types of expertise will find synergy points and attack the problem from their perspective to innovate in ways that are not possible when working in isolation.

\section{Conclusion}
\label{sec:conclusion}

The vision we have put forward is a synthesis of practical and research challenges for Technical Debt management. 
What is unique in the vision is that it challenges the research and practitioner communities that a shift in perspective is possible. It is possible to scope research problems to make progress on the most pervasive issues, extending the focus of Technical Debt research beyond code. While code-level artifacts can be accurately analyzed and measured, what really matters in large and evolving software systems is architecture Technical Debt and its management, but is more elusive and much harder upon which to collect data.

Technical Debt is a pervasive phenomenon in software systems. It always existed in systems, it will continue to exist in the future. A best path for improved Technical Debt management involves researchers, practitioners, and tool vendors working collaboratively and aggressively in parallel to develop tools, empower culture change to embrace practices, and develop Technical Debt management financial models to shift the focus on value generation to avoid lamp post research. 

\section*{Acknowledgment}
Copyright 2023 IEEE.
Authors Ozkaya and Shull's contributions to this material is based upon work funded and supported by the Department of Defense under Contract No. FA8702-15-D-0002 with Carnegie Mellon University for the operation of the Software Engineering Institute, a federally funded research and development center.
References herein to any specific commercial product, process, or service by trade name, trade mark, manufacturer, or otherwise, does not necessarily constitute or imply its endorsement, recommendation, or favoring by Carnegie Mellon University or its Software Engineering Institute.
DM23-0979

%
\bibliographystyle{IEEEtran}
\bibliography{bibliography.bib}

\end{document}